\documentclass{jaa}
\usepackage{natbib}
\usepackage[normalem]{ulem}
\usepackage{hyperref}
\hypersetup{
    colorlinks=true,
    linkcolor=blue,
    filecolor=magenta,      
    urlcolor=cyan,
    citecolor=blue,
    pdftitle={Overleaf Example},
    pdfpagemode=FullScreen, }

\usepackage{graphicx}
\usepackage{cancel}
\usepackage{subcaption}
\usepackage{stfloats}

\begin{document}\sloppy

\title{Impact of Galaxy Cluster Environment on the Stellar Mass Function of Galaxies}


\author{Sana Begum Murtuja Shaikh\textsuperscript{1,*} and Priyanka Singh\textsuperscript{1,*}}
\affilOne{\textsuperscript{1}Department of Astronomy, Astrophysics and Space Engineering, Indian Institute of Technology, Indore 453552, India.}


\twocolumn[{

\maketitle

\corres{ms2304121003@iiti.ac.in, psingh@iiti.ac.in}

\msinfo{}{}

\begin{abstract}
Galaxy clusters represent some of the most extreme environments in the Universe. They are ideal locations to study the impact of an extreme environment on the evolution of the Stellar Mass Function (SMF), which describes the statistical distribution of galaxies as a function of their stellar masses. In this work, we examine how the SMF of galaxies depends on factors such as the surrounding environments, whether they reside in isolated fields or clusters. We use the 9-band photometric galaxy data of the G9 patch from the Kilo Degree Survey (optical) and the VISTA Kilo-Degree Infrared Galaxy Survey (infrared), containing around 3.7 million galaxies, overlapping with the cluster catalog provided by the eROSITA Final Equatorial Depth Surveys (eFEDS). After applying appropriate selection criteria, we have 105 eFEDS clusters within the redshift range 0.385-0.8, covering $\sim 46$ square degrees. The large, continuous overlap of the surveys allows us to examine the SMF of the cluster galaxies within the cluster-centric radial bins up to $5R_{500}$. We find a clear detection of the cluster galaxy SMF up to $2R_{500}$ beyond which it's consistent with the background. We divide the cluster sample into redshift, mass, and X-ray luminosity bins to examine their impact on the SMF.  The SMF of cluster galaxies for the high-mass clusters shows a decline at low stellar masses ($M_*\lesssim  2\times 10^{10}M_\odot$) within $0-0.5R_{500}$, as compared to a flat SMF for the low-mass clusters, suggesting the low-mass galaxies grow over time before reaching the cluster center. Additionally, we find a flatter SMF for the low redshift bin within $0.5R_{500}$ at stellar masses  $M_*< 10^{10}M_\odot$. We also examined the effect of cluster ellipticity on the cluster galaxy SMF; however do not find statistically significant differences between the high and the low ellipticity clusters.
\end{abstract}

}]


\doinum{12.3456/s78910-011-012-3}
\artcitid{\#\#\#\#}
\volnum{000}
\year{0000}
\pgrange{1--}
\setcounter{page}{1}
\lp{1}

\section{Introduction}
Galaxies contain gas, dust, stars, and dark matter, bounded by their gravitational fields. The stellar properties of these galaxies depend heavily on whether they are in isolated environments, groups, or clusters. Galaxy clusters are the largest gravitationally bound structures that contain hundreds to thousands of galaxies. Galaxy clusters go through various environmental processes such as tidal interactions due to either galaxy harassment or interaction with global cluster potential, tidal forces (\citealt{BialasrefId0}; \citealt{Merritt1984ApJ...276...26M}; \citealt{Byrd1990ApJ...350...89B}; \citealt{Gnedin2003ApJ...589..752G}), ram pressure stripping due to the presence of hot ($T \sim 10^7-10^8 K$) intracluster medium (ICM) (\citealt{Gunn1972ApJ...176....1G}; \citealt{Jaff2015MNRAS.448.1715J}), and galaxy mergers, which are more common in groups than in clusters due to the high velocity dispersion of cluster galaxies (\citealt{makino1997ApJ...481...83M}; \citealt{Pearson_2024}). These processes generally reduce galaxies' star formation rate, affecting their evolution, while field galaxies residing in less dense and isolated regions undergo fewer interactions with neighboring galaxies, resulting in a higher star formation rate than cluster galaxies. 

Photometric surveys such as the Dark Energy Survey (DES; \citealt{Abbott_2018}), the Two Micron All Sky Survey (2MASS; \citealt{Skrutskie_2006}), the COSMOS/UltraVISTA Survey (\citealt{Muzzin_2013}), the Kilo Degree Survey (KiDS; \citealt{KUijken2015MNRAS.454.3500K}; \citealt{dejong2015refId0}), the VISTA Kilo-Degree Infrared Galaxy Survey (VIKING; \citealt{Edge2013Msngr.154...32E}), etc., enable the mapping of large-scale structures in the universe such as galaxies, galaxy clusters, and many more. Studying the observational quantities, such as the stellar mass functions (SMFs) as a function of cluster-centric radii with the help of such large sky photometric surveys, can help us uncover where exactly galaxies start feeling the influence of the cluster's environment. Comparing the SMFs of galaxies in different environments and at different redshift ranges can allow us to uncover valuable insights into the process that shapes galaxy evolution.

Several observational studies have tried to quantify the impact of the environment on the stellar mass function of galaxies (e.g., \cite{vanderburg2013A&A...557A..15V, Nantais2016A&A...592A.161N, Tomczak2017MNRAS.472.3512T, van2018stellar, Kim_2023}). In \cite{van2018stellar}, the authors studied 21 of the most massive clusters identified through the Planck SZ survey with multiband photometry ranging from the u band ($\sim 3,000\, \rm A^{\circ}$) to the Ks ($\sim 24,000\, \rm A^{\circ}$) band with the redshift range 0.5 to 0.7, to understand how the environment affects galaxies as they move towards the cluster center. They compared the SMF of galaxies within cluster environments to that of galaxies in the field at a similar redshift bin, using the COSMOS/UltraVISTA survey's dataset. 
The SMF comparison of the field to cluster galaxies revealed a higher quenched fraction of galaxies in the clusters than the field. The shape of the SMF of the star-forming galaxies is the same for galaxy clusters and the field, thus independent of the environment. They found that the quenched fraction of low-mass galaxies is higher near the cluster center, and it decreases while moving away from the cluster center. The environment quenching causes a decrease in the quiescent galaxy SMF at lower stellar masses. The study did not find a significant change in the SMF between the radial bins. Note that their study is based on a small sample of massive galaxy clusters, with the coeval field galaxy sample based on a different photometric dataset covering a small fraction of the sky. Using a larger sample of galaxy clusters, \cite{Kim_2023} studied 1,626 galaxies with spectroscopic observations in 105 galaxy clusters in the redshift bin (0.26, 1.13) to quantify the effect of the cluster environment on the star formation of their galaxies. They found that the longer the galaxy stays in the cluster, the older its stellar population becomes, suggesting the slowdown of star formation due to environmental quenching, irrespective of the redshifts and galaxy luminosities. However, such spectroscopic studies are limited by small galaxy samples.

In this work, we use the KiDS+VIKING-450 (KV450; \citealt{Wright2019A&A...632A..34W}) galaxy dataset for galaxy properties (e.g., stellar masses and photometric redshifts) estimated using the 9-band (\textit{ugriZYJHKs} with wavelength ranging from 3,000 to 24,000 \AA) photometry and the galaxy cluster sample with a total mass ranging from $10^{13.17} M_\odot$ to $10^{15.08}M_\odot$ distributed over a redshift range of 0.01 to 1.3 from the eROSITA Final Equatorial-Depth Survey (eFEDS; \citealt{BrunnerrefId0}; \citealt{Liu2022A&A...661A...2L}) to investigate the effect of environment on galaxies. The continuous, wide sky coverage of these multi-wavelength surveys provides us a large sample of galaxies in different environments for a comprehensive analysis. We adopt $\Lambda$CDM cosmology where $\Omega_{\text{m}} = 0.3$, $\Omega_{\Lambda} = 0.7$, $H_0 = 70 \, \text{km s}^{-1} \, \text{Mpc}^{-1}$, $\Omega_{\text{b}} = 0.022/h^2_0$, $\sigma_8=0.8$, and $n_s=0.96$.

\section{Datasets}

\begin{figure*}
    \centering
    \includegraphics[width = 15cm]{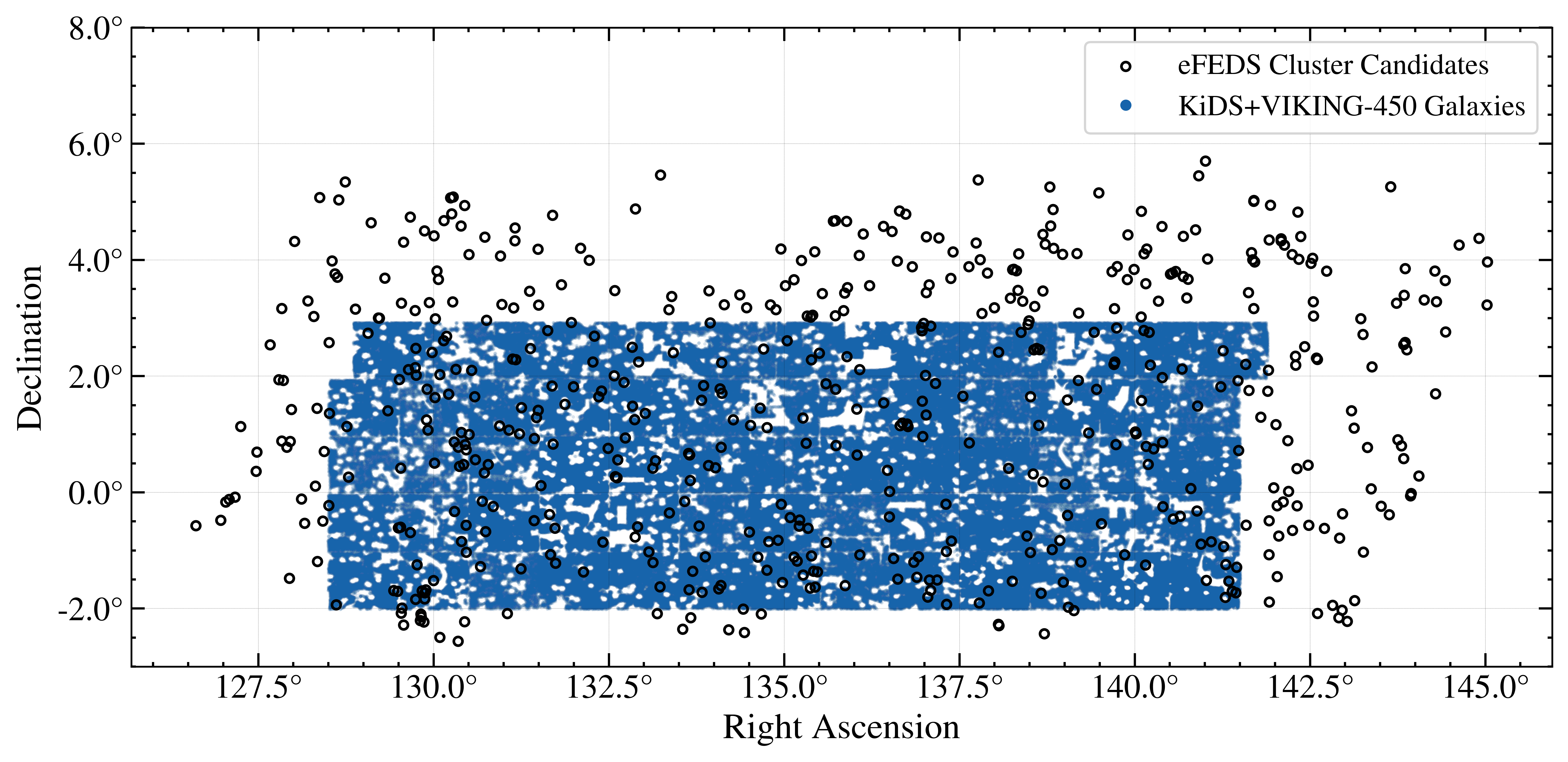}
    \caption{The overlapping sky region between the KV450 galaxy catalog and the eFEDS cluster catalog. The blue-colored region shows the final sample of KV450 galaxies in the G9 patch, and the black circles show eFEDS clusters.}
    \label{fig:Efeds field}
\end{figure*}

\subsection{KiDS+VIKING-450 Galaxy Sample}
\label{sec:KV450}

The KiDS and VIKING surveys are designed to provide complementary optical and near-infrared (NIR) wavelength coverage scanning $\approx 1350$ deg\textsuperscript{2} of the sky in the Galactic North and the South regions. Both surveys have already completed their imaging over the planned sky area, VIKING with DR4 in February 2020 and KiDS with DR5 in June 2024 (\citealt{Wright2024A&A...686A.170W}). \cite{Wright2019A&A...632A..34W} combined optical \textit{ugri} bands (3,000-9,000 \AA) from the KiDS survey and near-infrared \textit{ZYJHKs} bands (8,000-24,000 \AA) from the VIKING survey using the 450 deg\textsuperscript{2} overlapping coverage area of these surveys. The combined KV450 sample is defined as only those KiDS-450 sources that overlap with VIKING imaging and mask all other sources with near-infrared coverage missing with a depth $\leq 25$ in the r-band. The use of a nine-band wavelength coverage enables substantial reductions in systematic uncertainties, including outlier rates and scatter, when comparing photometric and spectroscopic redshifts. The catalog presented by \cite{Wright2019A&A...632A..34W} covers $\sim341$ deg\textsuperscript{2} with 447 overlapping pointings, consisting of $\sim49$ million sources, including stars and galaxies with nine-band photometry, estimated photometric redshifts, and stellar masses for the sources. We remove stars as well as galaxies for which the stellar masses were undefined, thus reducing the sample to $\sim31.45$ million galaxies.

\subsection{eFEDS Clusters}
\label{eFEDS Clusters}
eFEDS was conducted during the performance evaluation stage of the eROSITA X-ray telescope in the first quarter of November 2019. It is designed to serve as proof of concept for the eROSITA All-Sky Survey (eRASS; \citealt{MerlonirefId0}) spanning $ \sim $140 deg\textsuperscript{2} ($126^{\circ} < \text{RA} < 146^{\circ} \text{ and } -3^{\circ} < \text{Dec} < +6^{\circ}$) of the sky. \cite{Liu2022A&A...661A...2L} presented the catalog of 542 candidates for groups and galaxy clusters identified in the eFEDS field. These clusters lie in the redshift range from 0.01 to 1.3, with a median redshift of 0.35 and a likelihood extent larger than 6, leading to an 80\% purity of the sample. They used the multi-component matched filter (MCMF) cluster confirmation tool (\citealt{Klein2018MNRAS.474.3324K}) to compute photometric redshifts for the clusters by analyzing the photometric dataset (from DECaLS, unWISE, and HSC surveys) and finding the galaxy overdensities along the cluster positions. Spectroscopic redshifts were derived by cross-matching with various publicly available spectroscopic surveys, such as the 2MASS Redshift Survey (2MRS; \citealt{Huchra_2012}), the Sloan Digital Sky Survey (SDSS; \citealt{Blanton2017AJ....154...28B}) up to DR16, and the Galaxy And Mass Assembly (GAMA; \citealt{Driver2011MNRAS.413..971D}) redshift survey. They provided spectroscopic redshifts for a total of 297 clusters.

\cite{Klein2022A&A...661A...4K} performed the optical follow-up of the eFEDS group and cluster candidates mentioned above to provide their redshift, dynamical state, richness, and masses. The mass and redshift distribution of these clusters are shown in the left-hand and right-hand panels of Figure \ref{fig: mass redshift distribution}, respectively. 

\subsection{Cluster Galaxy Sample}

\begin{figure*}
        \includegraphics[height=6.5cm,angle=0.0 ]{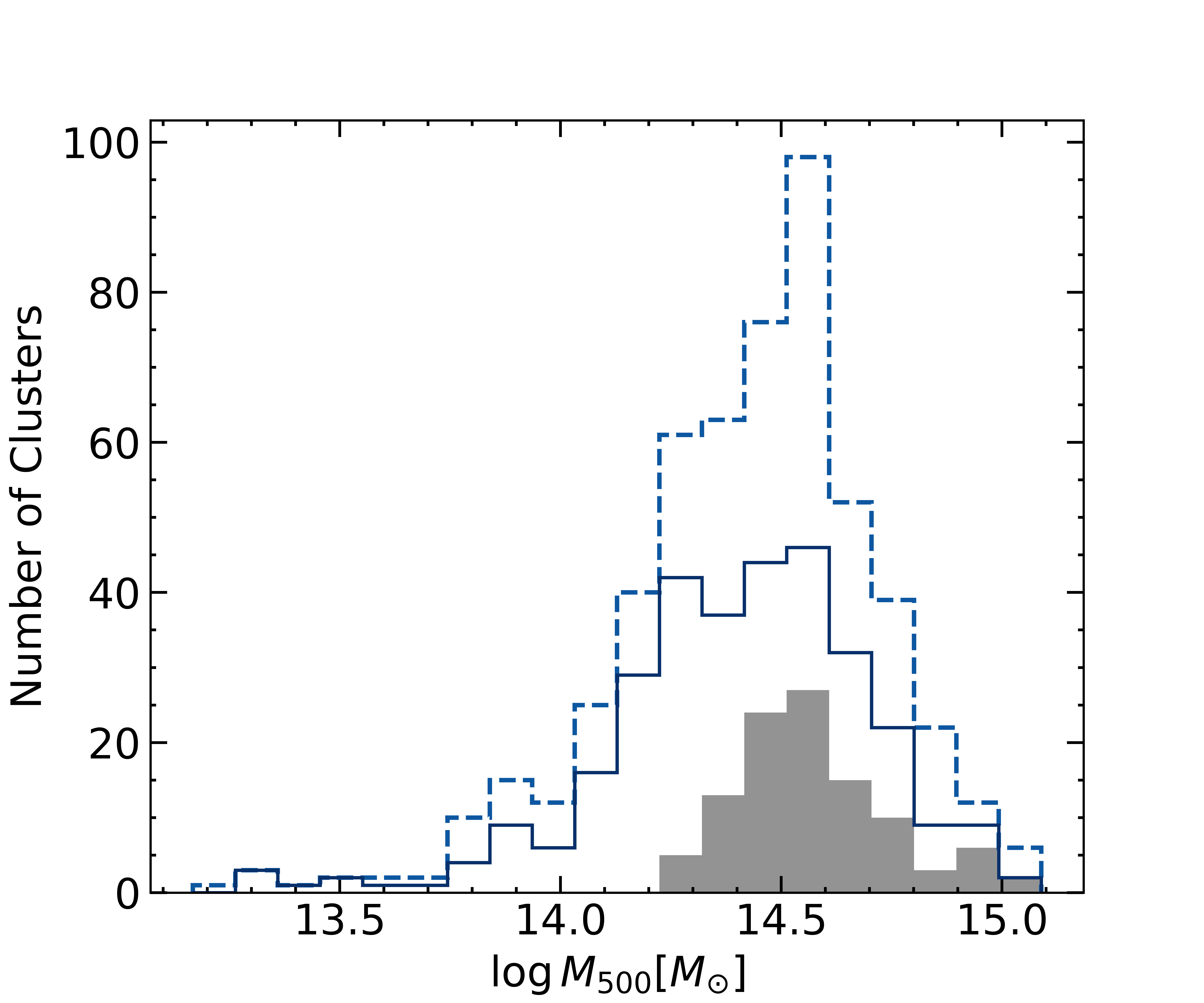}
           \hspace{0.05\textwidth}
        \includegraphics[height=6.5cm,angle=0.0 ]{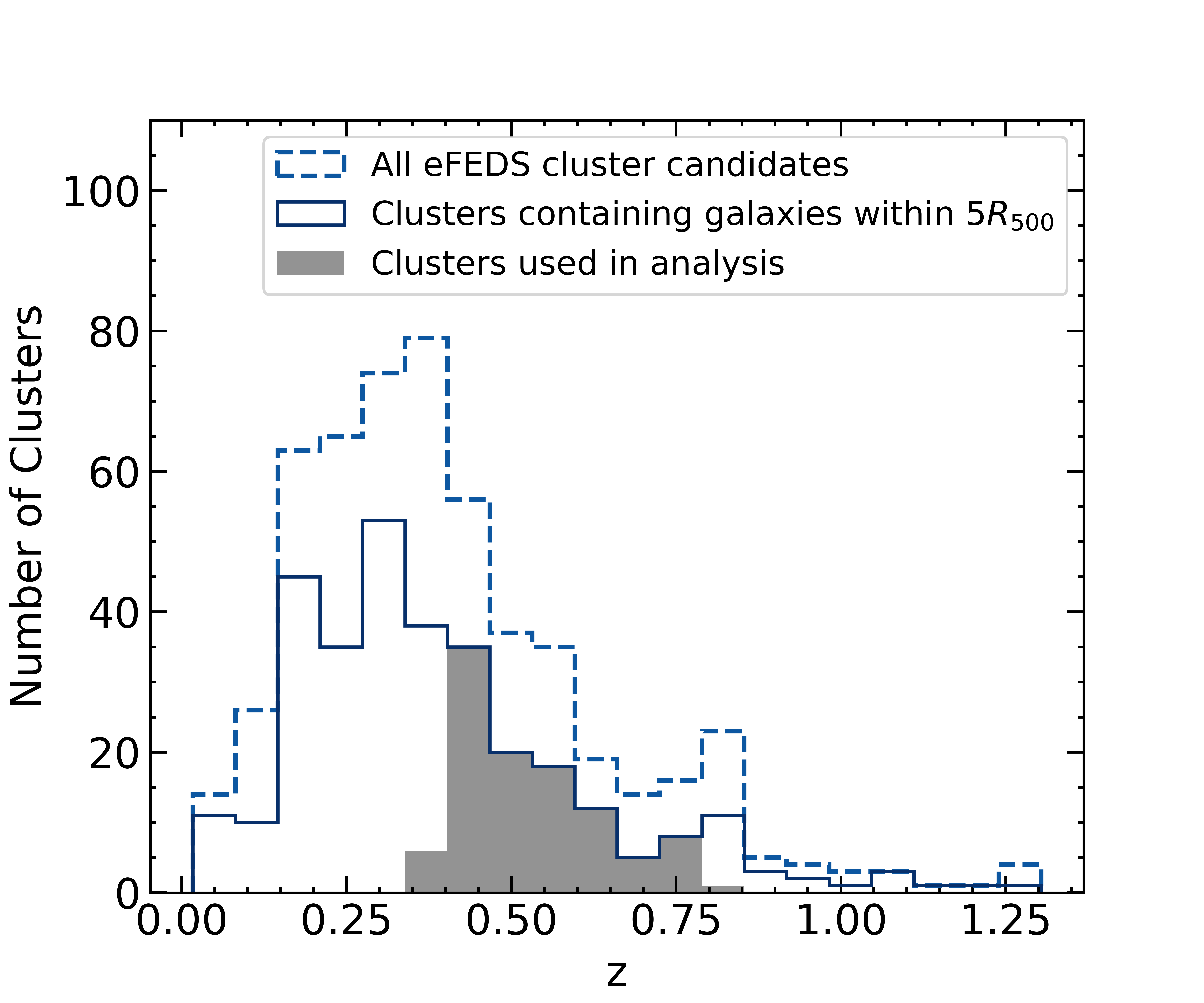}
     \caption{ Total mass (left-hand panel) and redshift (right-hand panel) distribution of eFEDS clusters. The dashed blue line shows all 542 clusters, the solid black line represents the 313 clusters that have KV450 galaxy coverage within $5R_{500}$, and the grey-shaded region shows the final sample of 105 clusters used in SMF analysis within the redshift range (0.385, 0.8).}
    \label{fig: mass redshift distribution}  
\end{figure*}

 The KiDS-450 dataset is divided into five distinct patches on the sky named G15, G12, G9, GS, and G23, primarily centered on fields observed by the GAMA redshift survey except GS (\citealt{Wright2019A&A...632A..34W}). The eFEDS field overlaps with the G9 patch of KV450 shown in Figure \ref{fig:Efeds field}. From the overlap, we created a cluster galaxy sample using KV450 galaxies within five times the eFEDS cluster boundary and the cluster redshift range. The cluster boundary is defined using $R_{500}$, where the average density within the cluster boundary is 500 times the critical density of the universe. We have calculated $R_{500}$ for a given $M_{500}$ for each cluster by using the relation:
\begin{equation}
M_{500} \equiv 500 \left( \frac{4\pi}{3} \right) R_{500}^3 \times \rho_c(z),
\end{equation}
where $\rho_c(z)$ is the critical density of the Universe at redshift $z$.

We select eFEDS clusters overlapping with the G9 patch within $5R_{500}$. This reduces the sample size to 313 clusters, shown by a solid black line in Figure \ref{fig: mass redshift distribution}.

The cluster redshift uncertainties provided by \cite{Klein2022A&A...661A...4K} are lower compared to the photometric redshift uncertainties of KV450 galaxies. Therefore, we apply redshift selection criteria using the photometric redshift uncertainties of galaxies estimated by \cite{Wright2019A&A...632A..34W} from 9-band photometry for the KV450 using Bayesian Photometric Redshifts (BPZ; \citealt{Benitez2000ApJ...536..571B}). They compare spectroscopic redshifts ($z_{\text{spec}}$) using a spectroscopic redshift sample from various overlapping surveys such as zCOSMOS (\citealt{Lilly2007ApJS..172...70L}), VIMOS
VLT Deep Survey (\citealt{Fvre2013TheVV}), DEEP2 Redshift Survey (\citealt{Newman2013ApJS..208....5N}), etc., to the estimated photometric redshifts ($z_{\text{ph}}$):
\begin{equation}
\frac{{z_{\text{ph}} - z_{\text{spec}}}}{{1 + z_{\text{spec}}}} \equiv \frac{\Delta z}{{1 + z}}
\end{equation}
\begin{equation}
\frac{\Delta z}{1 + z} = \sigma_m
\label{sigmam}
\end{equation}
where $\sigma_m$ is the median absolute deviation. The values for $\sigma_m$ are 0.061 for $z_{\text{ph}}\leq 0.9$ and 0.096 for $z_{\text{ph}} > 0.9$.

\section{Stellar Mass Function}

The SMF is defined as the number density of galaxies as a function of their stellar masses. In this paper, we calculate the SMF for the KV450 galaxies as a function of their host cluster mass, X-ray luminosity, distance from the cluster center, and redshift to understand the impact of the cluster environment on galactic evolution.

To construct the SMF, we first apply the stellar mass completeness limit on the galaxy sample at the higher redshift limit within the redshift range under consideration. The mass completeness limit is the minimum stellar mass of a galaxy at a specific redshift to be detected by a given survey. Galaxies with stellar masses below this limit are too faint to be detected accurately by the survey instruments, and many of them get missed, causing incompleteness while analyzing galaxy samples. We use the stellar mass completeness limit given by \cite{Wright2019A&A...632A..34W} (see their Figure 13), calculated as flux scale-corrected stellar masses using the LE PHARE (\citealt{Arnouts1999MNRAS.310..540A}; \citealt{IlbertO2006A&A...457..841I}) template fitting code. We use stellar-mass bins of bin width 0.25 dex to compute the SMF.

As we apply masks (e.g., bright foreground stars, incomplete photometric coverage) on the KV450 data set, we start to lose some of the coverage area of the survey. In addition, the clusters near the survey boundary are not evenly covered. Therefore, we need to take into account the missing areas in the cluster regions while constructing the SMF. To do this, we create a sky coverage map for the KV450 galaxy dataset using Hierarchical Equal Area isoLatitude Pixelization (HEALPix; \citealt{2005ApJ...622..759G}), which divides the celestial sphere into equal area pixels. We use the healpy (\citealt{Zonca2019}) Python library developed to handle pixelated data on a sphere based on the HEALPix scheme. We use $\rm N_{side}=4096$, which corresponds to a pixel resolution equal to $\approx 52$ arcseconds. 

The average number of galaxies per pixel is 16 in the KV450 footprint with a pixel resolution of $52$ arcseconds. To further avoid shot noise due to the low number of galaxies per pixel, we remove pixels that contain less than five galaxies. After applying these selection criteria, we get a final usable survey area for KV450 $\approx 379.67$ deg\textsuperscript{2}, and the usable area for the G9 patch is $\approx 46$ deg\textsuperscript{2}.

\begin{figure}
    \centering
    \includegraphics[width = 0.9\linewidth]{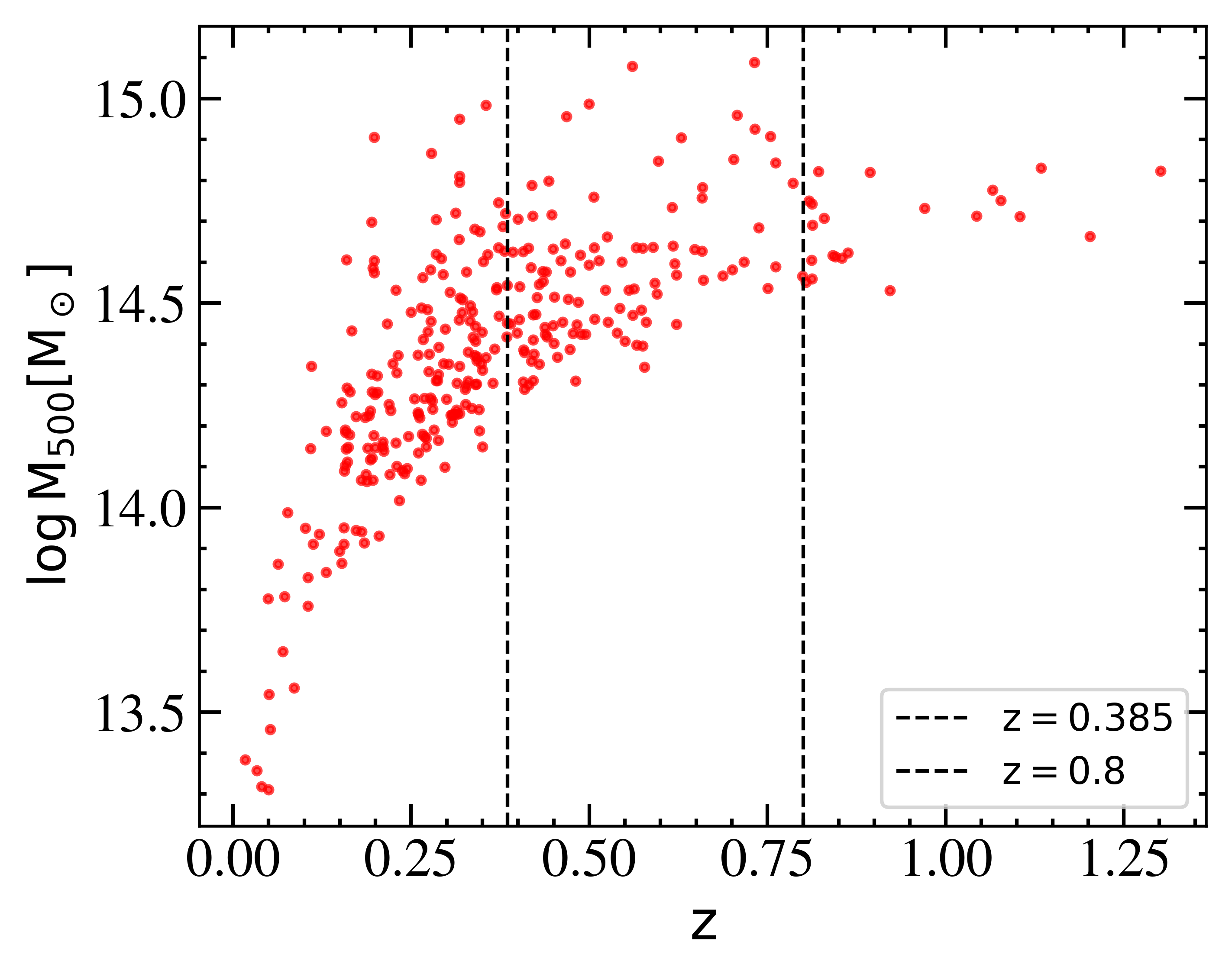}
    \caption{The distribution of the total mass for 313 eFEDS clusters (red filled circles) as a function of their redshifts. The final sample of 105 eFEDS clusters used for SMF analysis lies between z = (0.385, 0.8), shown by the dashed vertical lines.}
    \label{fig:mass vs redshift}
\end{figure}

\begin{figure*}
    \centering
    \includegraphics[width = 15cm]{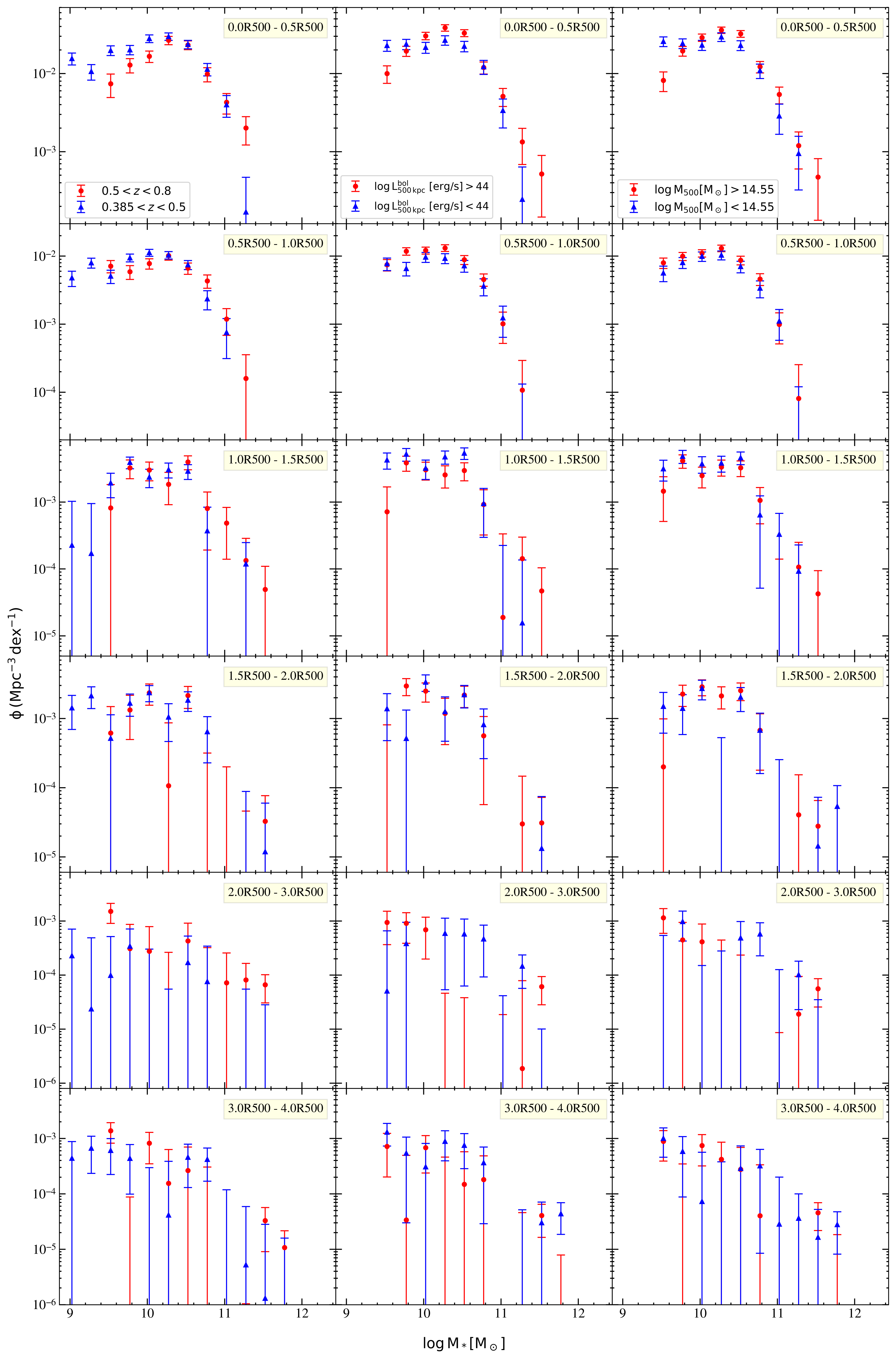}
    \caption{\textbf{Left-hand panel:} The comparison of the SMF of cluster galaxies in the redshift bins (0.385, 0.5) and (0.5, 0.8) for six cluster-centric radial bins. \textbf{Middle panel:} The SMF comparison for the cluster X-ray luminosity bins ($\rm 10^{43.65}, 10^{44}$) $\rm erg\, s^{-1}$ and ($\rm 10^{44}, 10^{45.1}$) $\rm erg\, s^{-1}$ for the entire redshift range (0.385, 0.8). \textbf{Right-hand panel:} The SMF comparison for the cluster mass bins ($\rm 10^{14.27}, 10^{14.55}$) $M_{\odot}$ and ($\rm 10^{14.55}, 10^{15.08}$) $M_{\odot}$ for the entire redshift range (0.385, 0.8).}
    \label{fig:All type SMF}
\end{figure*}

\subsection{SMF of cluster galaxies}
To investigate the impact of the cluster's influence on the stellar properties of galaxies, we construct the SMF as a function of distance to the cluster center. To do this, we divide the region around the cluster into radial bins of size $0.5R_{500}$ out to $2R_{500}$. Beyond that, we increase the radial bin size as the SMF approaches the background level. We analyze the clusters out to $5R_{500}$.

The SMF of cluster galaxies in the $i^{th}$ radial bin and stellar mass bin centered around $M_*$ is given by,

\begin{equation}
\phi_i(M_*) = \frac{1}{\Delta \log M_*} \times \frac{N^g_i}{\sum_j V^c_{j,i}}
\end{equation}
and the uncertainty in the SMF due to the Poisson uncertainty in the galaxy number counts is given by,

\begin{equation}
\Delta \phi_i(M_*) = \frac{1}{\Delta \log M_*} \times \frac{\sqrt{N^g_i}}{\sum_j V^c_{j,i}}
\end{equation} 
where $\Delta \log M_*$ is the stellar mass bin width (0.25 dex), $N^g_i$ is the total number of galaxies (above the stellar mass completeness limit corresponding to the highest redshift cluster considered) in the $i^{th}$ radial bin of all the clusters considered, and $V^c_{j,i}$ is the volume of the $i^{th}$ radial bin of the $j^{th}$ cluster. 

To compute $V^c_{j,i}$, we considered the projection of the cluster over the sky as a cylinder with height as a comoving length associated with $z\pm \Delta z$ (where $\Delta z$ is the uncertainty in the cluster's galaxies, calculated using equation \ref{sigmam}) is given by 

\begin{equation}
V^c_{j,i} = A^c_{j,i} \times \left[ D_c(z + \Delta z) - D_c(z - \Delta z) \right]
\end{equation}
where $A^c_{j,i}$ is the area of the $i^{th}$ radial bin of the $j^{th}$ cluster (after masking), and $D_c$ is the comoving distance.

\subsection{Background/Foreground Subtraction}
We use photometric redshift uncertainty $\Delta z$ to assign galaxies to the host cluster, as a result of which we also include background and foreground galaxies with large redshift uncertainties that are not part of the cluster. To overcome this, we perform background/foreground subtraction by utilizing galaxies in the radial range $4R_{500} - 5R_{500}$, far from the cluster, to ensure that there is no influence of the cluster on their SMF. We apply the same redshift selection as in the case of cluster galaxies while masking the clusters out to a radius of 2$R_{500}$ to compute the background level of the SMF. Note that we calculate the background SMF using only those cluster outskirts for which the cluster galaxy SMF is calculated, while dividing the clusters based on their redshifts and masses. (see Sections \ref{sec-z-dep} and \ref{sec-mass-dep}). We then subtract the background level from the inner cluster radial bins to get the final SMF of the cluster galaxies.

\section{Results}

\begin{table*}
\centering
\renewcommand{\arraystretch}{1.8}
\resizebox{0.95\textwidth}{!}{
\begin{tabular}{|c|c|c|c|c|}
\hline
\textbf{\Large$\phi(M_*)$} & \multicolumn{2}{c|}{\textbf{\large$\rm0.0 R_{500} - 0.5 R_{500}$}} & \multicolumn{2}{c|}{\textbf{\large$\rm0.5 R_{500} - 1.0 R_{500}$}} \\
\cline{2-5}
             & $\textbf{$ \rm M_* < 2\times 10^{10} M_{\odot}$}$ & \textbf{$ \rm M_* > 2\times 10^{10} M_{\odot}$}& \textbf{$ \rm M_* < 2\times 10^{10} M_{\odot}$} &\textbf{$ \rm M_* > 2\times 10^{10} M_{\odot}$}\\
\hline
$0.385<z<0.5$  & $0.29 \pm 0.06$ &$-1.45 \pm 0.25$  &$0.23 \pm 0.07$  & $-1.99 \pm 0.42$ \\
\hline
$0.5<z<0.8$ & $0.70 \pm 0.16$ &$-1.47 \pm 0.25$  & $0.25 \pm 0.14$  & $-1.24 \pm 0.35$ \\
\hline
$\rm \log L^{\rm bol}_{\rm 500\, kpc} [\rm erg\, s^{-1}] < 44$       &$0.07 \pm 0.12$  &$-1.41 \pm 0.31$  &$0.16 \pm 0.15$  &$-1.42 \pm 0.40$  \\
\hline
$\rm \log L^{\rm bol}_{\rm 500\, kpc} [\rm erg\, s^{-1}] > 44$      & $0.67 \pm 0.12$  &$-1.66 \pm 0.22$  &$0.24 \pm 0.11$  &   $-1.52 \pm 0.34$\\
\hline
$\rm \log{M_{500} [M_\odot] < 14.55}$     &  $0.07 \pm 0.10$ &$-1.59 \pm 0.31$  & $0.31 \pm 0.15
$ & $-1.49 \pm 0.39$ \\
\hline
$\rm \log{M_{500} [M_\odot] > 14.55}$      & $0.65 \pm 0.12$ & $-1.59 \pm 0.20$ & $0.27 \pm 0.11$ & $-1.48 \pm 0.33$ \\
\hline
\end{tabular}}
\caption{Comparison of the slopes and their uncertainties for the power law fits derived at low stellar masses ($ M_* < 2\times 10^{10} M_{\odot}$) and high stellar masses ($M_* \sim 2\times 10^{10}-1.1\times 10^{11} M_{\odot}$) for the redshift, X-ray luminosity, and cluster mass dependent SMF distributions shown in Figure \ref{fig:All type SMF} for the two radial bins out to $R_{500}$. 
}
\label{tab:main_all_panels}
\end{table*}

Figure \ref{fig:mass vs redshift} shows the redshift-mass distribution of the 313 eFEDS clusters with KV450 overlap. We further exclude clusters below redshift 0.385, as the KV450 galaxy SMF below $z = 0.3$ had a significant deviation from the expected SMF due to the survey bias against selecting galaxies with the largest angular sizes (\citealt{Wright2018MNRAS.480.3491W, Wright2019A&A...632A..34W}) and the maximum value of $\Delta z$ in the cluster redshift bin (0.3, 0.4) is $\approx 0.085$. This ensures that we do not include any galaxies below redshift 0.3 while considering the photometric redshift uncertainties. We also apply an upper redshift limit at 0.8 since we are only left with the most massive galaxies due to the high stellar mass completeness limit at higher redshifts. The final usable size of the eFEDS cluster sample enclosed by these limits (dashed vertical lines in Figure \ref{fig:mass vs redshift}) is 105, with the median redshift of 0.5 and the median mass of $10^{14.55} M_\odot$.

\subsection{Redshift dependence}
\label{sec-z-dep}

We divide the cluster sample into two redshift ranges (0.385, 0.5) and (0.5, 0.8), where 0.5 is the median redshift of the final cluster sample. The stellar mass completeness limit is $9.2\times10^8 M_{\odot}$ and $2.7\times 10^9 M_{\odot}$ for the low redshift and high redshift samples, respectively. The left-hand panel of Figure \ref{fig:All type SMF} shows the cluster galaxy SMF as a function of redshift for different cluster-centric radial bins. 

We find a clear detection of cluster galaxy SMF out to $R_{500}$ throughout the stellar mass range considered, declining as we move to larger cluster-centric radii. The SMF is consistent with the background level beyond $2R_{500}$. The SMF peaks around $M_* \approx 1-2\times 10^{10} M_{\odot}$, a feature that is particularly prominent in the innermost radial bin $0-0.5R_{500}$. The decline in the SMF below masses $M_*<10^{10} M_{\odot}$ at $<0.5 R_{500}$ highlights that low-mass galaxies have low abundances near cluster centers, likely because such galaxies have low survival as they travel towards the cluster centers. We only find any differences in the SMF as a function of redshift in the innermost bin, at low stellar masses ($M_* <10^{10} M_{\odot}$). The SMF of lower-mass galaxies is higher at lower redshift. 

To further support our conclusions, we derive the slope of the SMF by dividing the sample into high and low stellar mass bins around $M_* \approx 2\times 10^{10} M_{\odot}$, assuming a power law relation between the SMF and the stellar masses. The corresponding slopes ($\alpha$) and their uncertainties are provided in Table \ref{tab:main_all_panels}. The slopes are derived only up to $R_{500}$ and $M_* < 1.1\times 10^{11}M_{*}$; beyond this, the increasing uncertainties in SMFs do not allow for performing the fitting analysis. We can clearly see that the low stellar mass end of the SMF is steeper in the inner-most radial bin at high redshifts (where $\alpha = 0.70\pm 0.16$) than the lower redshift bin (where $\alpha = 0.29\pm 0.06$), and it gets flatter if we move towards the outer radial bin, where the slopes for both the high and low redshift bins are consistent with each other. The low redshift-low stellar mass SMF slopes are in agreement between the radial bins $0-0.5R_{500}$ and $0.5-1R_{500}$ within $1\sigma$. The slopes for high stellar mass bins are independent of redshift within the uncertainty.

\begin{figure*}
    \centering
    \includegraphics[width = 13cm]{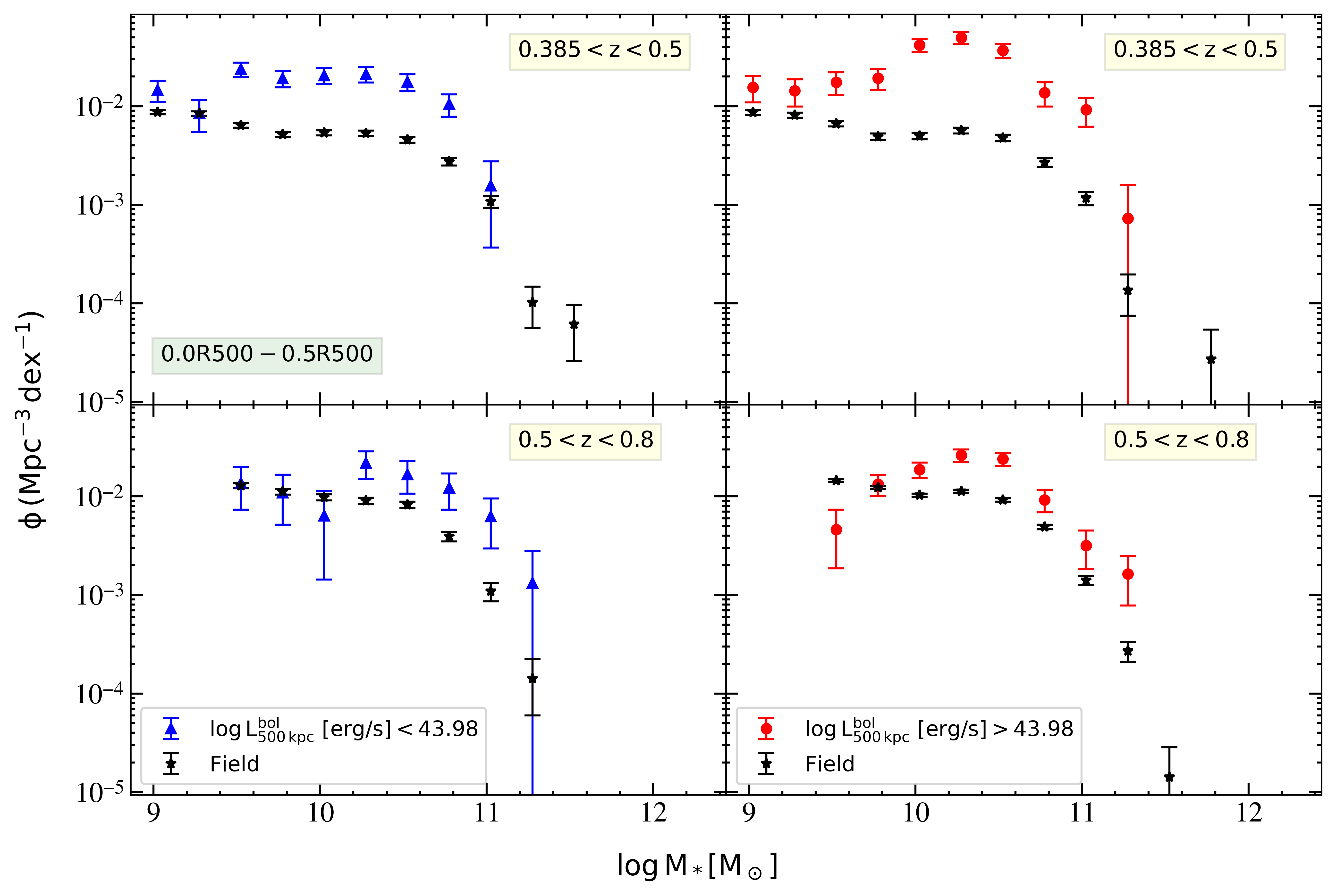}
    \caption{The comparison between the X-ray luminosity and redshift-dependent SMF of cluster galaxies within $0.5R_{500}$ of the cluster-centric radii and the coeval field galaxy SMF calculated within the cluster-centric radial bin $4-5R_{500}$. The top and the bottom panels correspond to low (0.385, 0.5) and high (0.5, 0.8) redshift bins, respectively. The blue triangles and red circles correspond to low ($\log L^{\rm bol}_{\rm 500\, kpc} [\rm erg\, s^{-1}] < 44$) and high ($\log L^{\rm bol}_{\rm 500\, kpc} [\rm erg\, s^{-1}] > 44$) X-ray luminosity bins, respectively, for both redshift bins. The coeval field galaxy SMF is shown by black stars in all the panels.}
    \label{fig:SMF lum field}
\end{figure*}

\subsection{Cluster mass and X-ray luminosity dependence}
\label{sec-mass-dep}
The X-ray luminosity of a galaxy cluster is a well-studied tracer of the total mass of the cluster, and the two show a tight scaling relation often used in cosmological studies (\citealt{Bulbul_2019, Singh2020}). In this section, we divide the SMF of cluster galaxies into X-ray luminosity ($L^{\rm bol}_{500\, \rm kpc}$) and halo mass ($M_{500}$) bins to study their impact on the stellar properties. The X-ray luminosity used here is the bolometric X-ray luminosity in the band (0.01-100) keV within the 500 kpc radius of the cluster. We used 96 clusters for which the luminosities were defined by \citealt{Liu2022A&A...661A...2L} for X-ray luminosity-dependent SMF calculations, and all 105 clusters are used for mass-dependent SMF. Since the X-ray luminosity and halo mass are correlated with each other, we expect to see similar trends in the SMF, and the aim here is to identify which quantity shows a clearer impact on the SMF.

We divide the 105 clusters on the basis of their median mass and X-ray luminosity. The median mass and X-ray luminosity of the sample are $10^{14.55} M_{\odot}$ and $10^{44} \rm erg/s$, respectively. The X-ray luminosity and mass dependence of the SMF are shown in the middle and right-hand panels of Figure  \ref{fig:All type SMF}, respectively. We show the comparison of the SMF slopes in different X-ray luminosity and cluster mass bins in Table \ref{tab:main_all_panels}. The slopes are consistent between X-ray luminosity and cluster mass-based division for the innermost radial bin. We find some differences in the outer radial bin at low stellar masses, which are still consistent within the uncertainties in the power law slopes.

We find a slightly clearer impact (though minor to draw any strong conclusions) of X-ray luminosity compared to the halo mass on the SMF. The SMF of high mass/X-ray luminosity clusters shows a peak around $M_* \approx 2\times 10^{10} M_{\odot}$, whereas the SMF flattens at low stellar masses for the low mass/X-ray luminosity clusters in the innermost radial bin (as shown in Table \ref{tab:main_all_panels} for $0-0.5R_{500}$, the low mass end slopes for high and low X-ray luminosity bins are $0.67\pm0.12$ and $0.07\pm0.12$, respectively). Note that the SMF of field galaxies continues to increase at lower stellar masses (\citealt{Muzzin2013ApJ...777...18M}). These results support that the SMF of low stellar mass galaxies is sensitive to their environment and that the abundances of these galaxies decline in overdense environments such as galaxy clusters.

We further divide the cluster sample into redshift and X-ray luminosity bins to disentangle the effect of redshift and cluster mass/X-ray luminosity on the SMF. In Figure \ref{fig:SMF lum field}, we show only the results for the innermost radial bin ($0-0.5R_{500}$), where we find the most interesting trends. Results for the outer radial bins are shown in \ref{appendixsec:lum_red_smf} We show the background levels, which can be treated as the coeval field levels around the cluster sub-samples. We clearly find the difference in the shape of the SMF between the high X-ray luminosity/high mass galaxy clusters and the coeval fields. The cluster SMF shows a peak around $M_* \approx 2\times 10^{10} M_{\odot}$ compared to the field SMF, which continues to increase at lower stellar masses. These results further highlight that the mass/X-ray luminosity dependence of the SMF is not due to any hidden redshift dependence, and it persists at both redshift bins considered here.

We also derived the slopes for the SMF-stellar mass power law relation for the results shown in Figure \ref{fig:SMF lum field}. At low stellar masses, the slope is steeper for the high X-ray luminosity-low redshift bin ($0.50 \pm 0.09$) than the low X-ray luminosity-low redshift bin ($0.13 \pm 0.09$), consistent with the trend found at high redshifts (with slopes $0.70 \pm 0.21$ and $0.32 \pm 0.30$ for high and low \textcolor{red}{X-ray} luminosity bins, respectively). The coeval field SMFs show a negative slope ($\approx -0.2$), i.e., the SMF increases with decreasing stellar masses, a feature in contrast with the cluster galaxy SMF at these stellar masses.

At high stellar masses, the slopes for cluster galaxy SMF and their coeval field SMF are approximately consistent with each other, irrespective of the luminosities and redshifts of these objects.

We qualitatively compare our results with \cite{van2018stellar}, which is closest to the redshift ($z \sim 0.5-0.7$), and the cluster mass ($M_{500} \sim 3\times 10^{14}-10^{15} M_{\odot}$) ranges to our final eFEDS cluster sample. In their Figure 4, they show the radial dependence of the cluster galaxy SMF out to $2R_{500}$. They do not find any significant radial dependence, though there is a minor decline in the total SMF (dominated by the SMF of quiescent galaxies) in the innermost radial bin at $M_* < 2\times 10^{10} M_{\odot}$ (also seen in \citealt{vanderburg2013A&A...557A..15V}). We find a stronger decline in the SMF, especially after dividing the sample into separate mass bins, thus highlighting that this feature is driven by the more massive end of our cluster sample. We are only able to do this thanks to the large cluster sample from the eFEDS survey with continuous photometric coverage from the KiDS and VIKING surveys.

\begin{figure*}

\centering

    \includegraphics[width = 18cm]{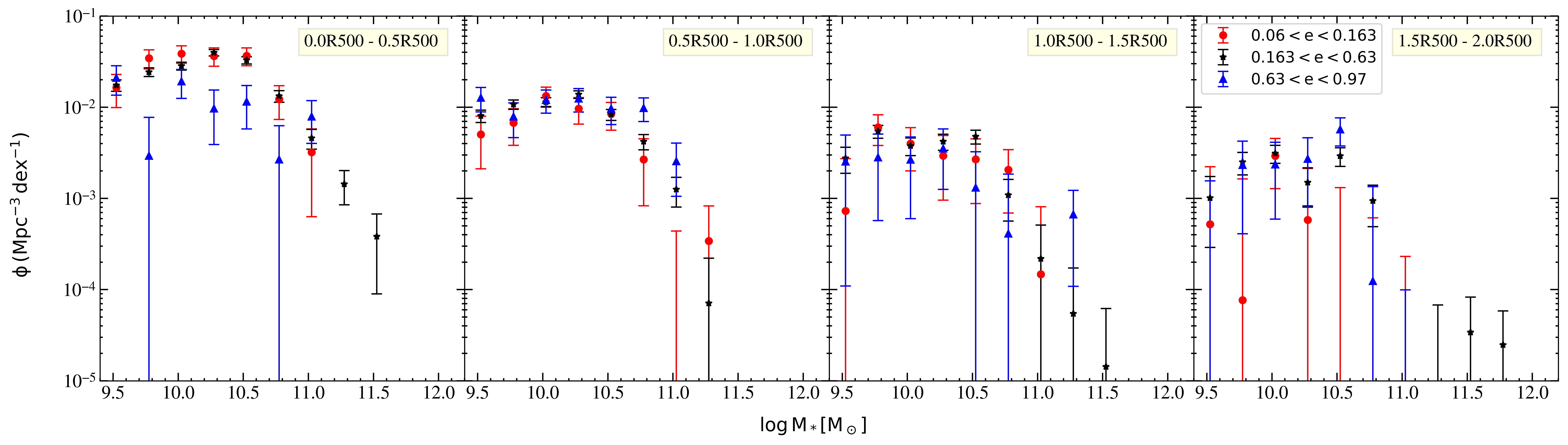}

    \caption{The comparison of SMF of cluster galaxies, based on the ellipticities of galaxy clusters up to $2R_{500}$ within the four radial bins. The blue triangles correspond to the average SMF of 10 clusters with the highest ellipticities, the red circles correspond to the average SMF of 10 clusters with the lowest ellipticities, and the black stars correspond to the remaining clusters with the intermediate ellipticities (mean level) from our final eFEDS cluster sample.}

    \label{fig:SMF_ellipticity}

\end{figure*}

\subsection{Dependence on dynamical state}
\label{sec_ellip_smf}
The dynamical state of a galaxy cluster is an indicator of the relaxation and merger history of the cluster. The stellar properties of the cluster galaxies are therefore connected to the dynamical state of their host. In this section, we examine the impact of the dynamical state of eFEDS clusters on the SMF of cluster galaxies. We use the ellipticity of the cluster as a proxy of the dynamical state of the cluster, where a lower ellipticity refers to more relaxed clusters and vice versa (\citealt{Gouin2021A&A...651A..56G, Liu2025ApJ...979..200L}). Also, the young clusters are less evolved and irregular and can be less elliptical.

In Figure \ref{fig:SMF_ellipticity}, we show the SMF of cluster galaxies within four radial bins out to $2R_{500}$ for ten of the eFEDS clusters with the highest (blue triangles) and the lowest ellipticities (red circles) from the final eFEDS sample of 105 galaxy clusters. We also show the mean level SMF with the clusters having intermediate ellipticities between the high and low ellipticity cluster samples. We found a cluster within the sample named eFEDSJ084246.9-000917, with redshift 0.415 and ellipticity 0.08, which is mentioned as an unrelaxed cluster in Table C.1 of \cite{Klein2022A&A...661A...4K}. We removed it from our lowest ellipticity cluster sample, ranging from (0.067 - 0.163), and then calculated the SMF. We found that galaxy clusters with low ellipticities have higher SMF compared to clusters with high ellipticities only within $0-0.5R_{500}$. We do not find any significant differences in the SMF as a function of ellipticity beyond $0.5R_{500}$, as the compared SMFs within other radial bins always coincide.

To quantify the significance of these results, we perform a binomial test to find the statistical significance of high/low ellipticity cluster SMF lying below/above the mean SMF level, respectively. The low ellipticity cluster-galaxy SMF is consistent with the mean level within the error bars; therefore, we do not run the test for these. We find a p-value $\sim 0.14$ for high ellipticity clusters; therefore, these results are inconclusive at the moment. Future studies based on a larger cluster sample could help in a more conclusive assertion of an ellipticity-based cluster-galaxy SMF analysis.


\section{Conclusions}
In this work, we investigate the impact of galaxy clusters on the stellar content of galaxies by estimating the SMF of cluster galaxies as well as their coeval field as a function of cluster redshift, mass, X-ray luminosity, and cluster-centric radii. To do this, we utilize the galaxy cluster sample from the eFEDS X-ray survey spanning a redshift range of $z \sim 0.39-0.8$ and a mass range of $M_{500} \sim 10^{14.27}-10^{15.08} M_{\odot}$, after applying appropriate selection criteria. We use the galaxy sample from KV450, i.e., the KiDS (optical) and VIKING (IR) survey combination, with an approximate overlap of 46 deg$^2$ with the eFEDS survey after removing the survey area contaminated by bright foreground stars and incomplete photometry. The stellar masses and photometric redshifts of galaxies are provided by \cite{Wright2019A&A...632A..34W}, where we use photometric redshift uncertainties of galaxies to select cluster members.

We construct a radially binned cluster galaxy SMF out to $5R_{500}$ and then divide the cluster sample into mass, redshift, X-ray luminosity, and ellipticity bins to further understand the impact of cluster properties on galactic evolution. The main findings of this work are summarized as follows.

\begin{itemize}

    \item The SMF of the cluster galaxies is detected out to $2R_{500}$. It decreases as we move away from the center of the cluster, and it is consistent with the background level beyond $2R_{500}$.

    \item The SMF shows a clear peak in the innermost radial bin ($0-0.5R_{500}$) around $M_* \approx 1-2\times 10^{10} M_{\odot}$. At the low-stellar mass end and high redshifts, the SMF is steeper (with an SMF-stellar mass power law slope of $\alpha \approx 0.7$) than at low redshifts ($\alpha \approx 0.3$). This feature is in contrast to the SMF of field galaxies, which continues to increase at lower masses (with a negative slope $\alpha \approx -0.2$). On further dissecting the cluster sample, we find that this trend is driven by the massive (and hence luminous) galaxy clusters (with a low-mass end slope of $\approx 0.67$), indicating that the survival of low-mass galaxies decreases with increasing cluster mass and decreasing distance from the cluster center. 

    \item The SMF of lower mass ($M_* <10^{10} M_{\odot}$) galaxies is slightly higher at lower redshifts at $r<0.5R_{500}$, beyond which we do not find any redshift-dependent trends. 

    \item The SMF of the highest ellipticity clusters is lower than the SMF of the lowest ellipticity clusters. These results may point towards a link between the dynamical age of the clusters and the buildup of stellar content of their galaxies. However, we fail to reject the null hypothesis using a binomial test; thus, the results are not statistically significant for the cluster-galaxy sample under consideration.

\end{itemize}

Our results highlight the influence of dense cluster environments, especially near the inner cluster region, on the buildup of the overall stellar mass of cluster galaxies. These results also highlight that it is the lower mass end of the SMF in the most massive clusters that are affected significantly. To further dissect the physical processes responsible for these trends, it would be important to classify galaxies into various categories (e.g., star-forming, quiescent) in future studies. A comparison with large hydrodynamical simulations with statistically significant cluster populations could also provide more insights into the underlying physical processes affecting cluster galaxy populations.


\section*{Acknowledgments}

We thank Eeshaan Beohar, Abhishek Darwai, Aman Dubey, Suman Majumdar, Abhinav Narayan, and Shubhi Tiwari for the helpful discussions and suggestions.
\vspace{-1em}



\bibliography{example} 

\appendix

\section{X-ray luminosity and redshift dependent SMF at $r>0.5R_{500}$}
\label{appendixsec:lum_red_smf}
In Figure \ref{fig:remaing Lum median Smf}, we show the SMF for X-ray luminosity-redshift sub-bins at $r>0.5R_{500}$, an extension of results shown in Figure \ref{fig:SMF lum field}. Figure \ref{fig:lum_red_0.5_1R500}, \ref{fig:lum_red_1_1.5R500}, \ref{fig:lum_red_1.5_2R500} show the cluster-galaxy and coeval field SMF comparison within $0.5-1R_{500}$, $1-1.5R_{500}$, $1.5-2R_{500}$, respectively. In Figure \ref{fig:lum_red_0.5_1R500}, we see a higher cluster galaxy SMF than the coeval field only in the high X-ray luminosity-low redshift bin in the stellar mass range $8\times10^{9} - 6\times10^{10} M_{\odot}$. In other cases, the cluster-galaxy SMFs are mostly consistent with coeval fields given the uncertainty on the SMFs. At $r>R_{500}$, the cluster galaxy SMFs are getting progressively lower than their coeval field SMFs as we move away from the cluster center, which is the effect of background/foreground subtraction within our SMF analysis.

\renewcommand{\thefigure}{A.\arabic{figure}}
\setcounter{figure}{0}
\begin{figure*}[htbp]
    \centering

    \begin{subfigure}[b]{0.6\textwidth}
        \centering
        \includegraphics[width=\linewidth]{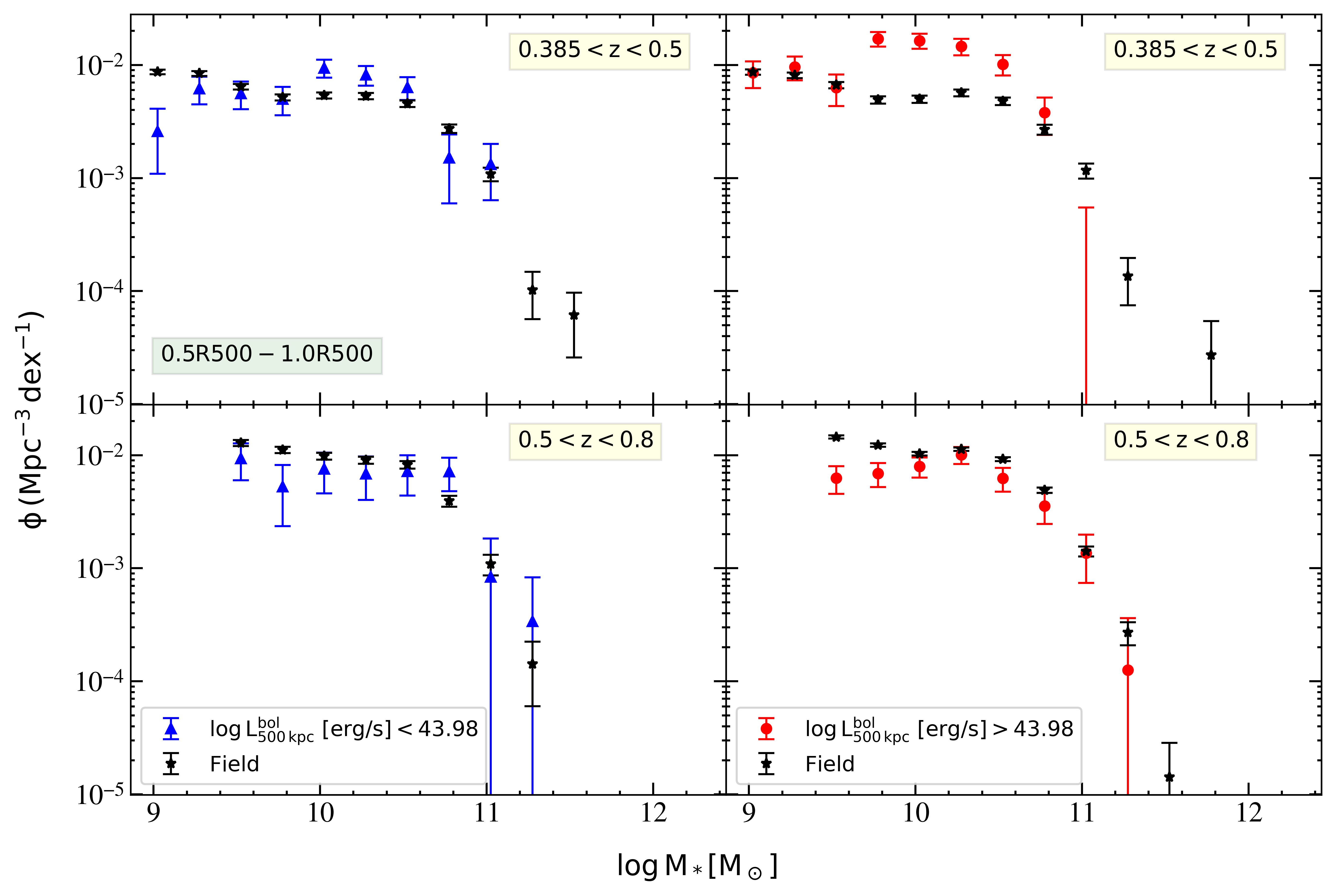}
        \caption{}
        \label{fig:lum_red_0.5_1R500}
    \end{subfigure}

    \begin{subfigure}[b]{0.6\textwidth}
        \centering
        \includegraphics[width=\linewidth]{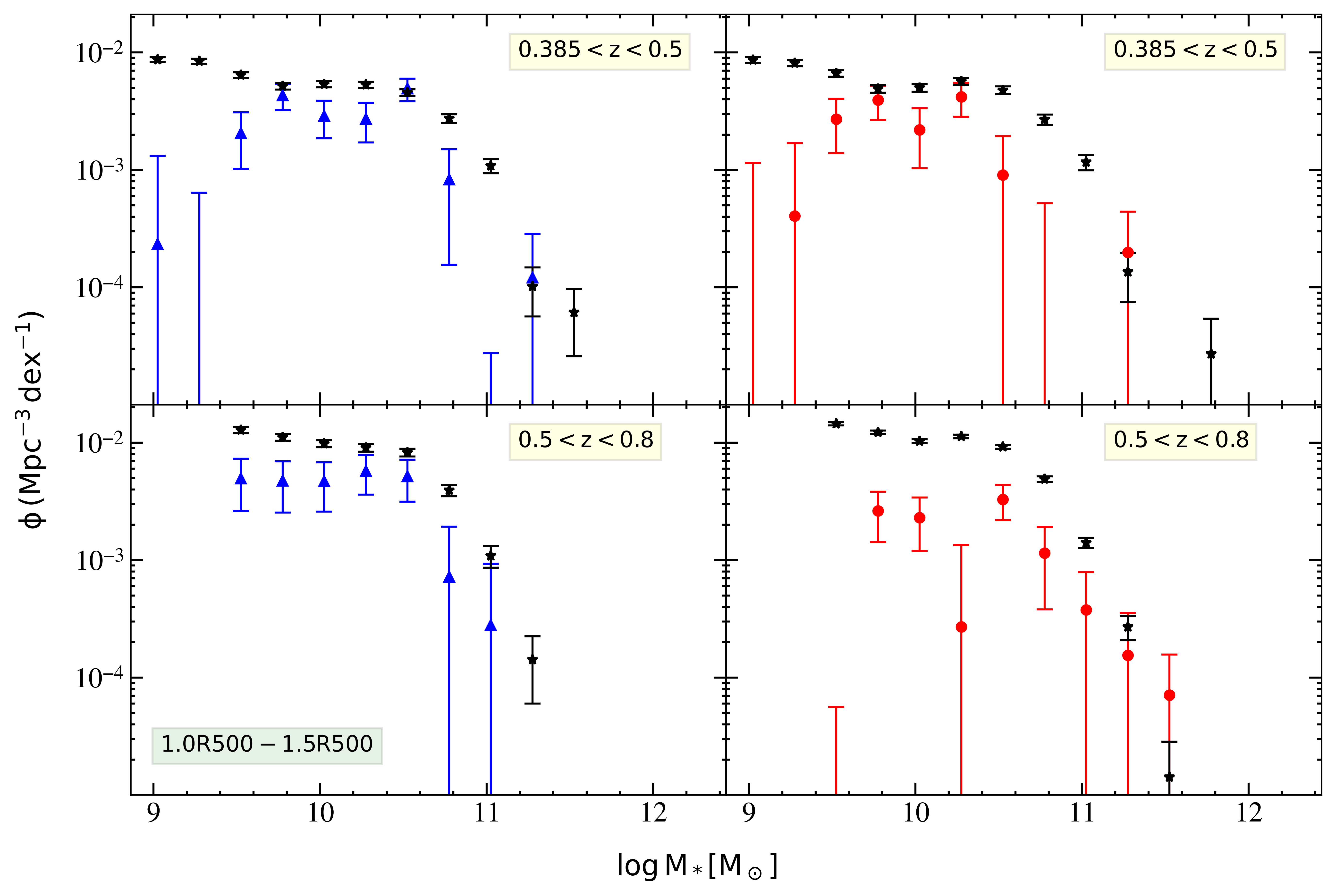}
        \caption{}
        \label{fig:lum_red_1_1.5R500}
    \end{subfigure}
    \hfill
    \begin{subfigure}[b]{0.6\textwidth}
        \centering
        \includegraphics[width=\linewidth]{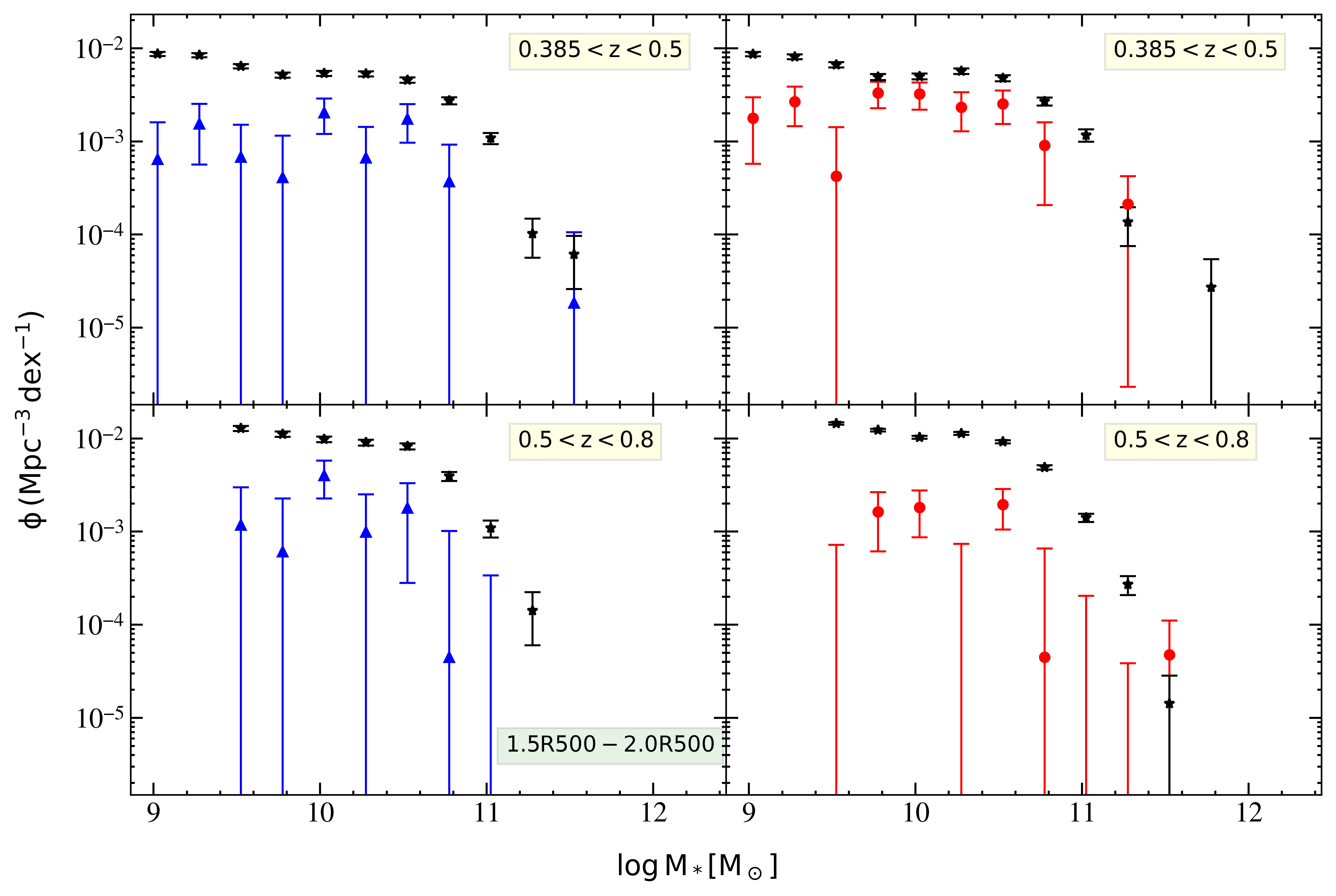}
        \caption{}
        \label{fig:lum_red_1.5_2R500}
    \end{subfigure}

    \caption{Same as Figure \ref{fig:SMF lum field} but for radial bins $0.5R_{500}-R_{500}$ (panel a), $R_{500}-1.5R_{500}$ (panel b), and $1.5R_{500}-2R_{500}$ (panel c).}
    \label{fig:remaing Lum median Smf}
\end{figure*}

\section{Trends of Ellipticity with Cluster Mass and Redshift}
\label{appendixsec:ellipticity_mass_redshift}
To investigate any biases due to the correlation between the ellipticity of the clusters and their masses/redshifts, we show the cluster's mass and redshift distribution as a function of cluster ellipticity for the final eFEDS cluster sample used in section \ref{sec_ellip_smf} in Figure \ref{fig:ellip_mass_redshift_distribution}. We do not find any significant correlation between the ellipticities and the clusters' masses or redshifts within the eFEDS cluster sample.
\newpage
\renewcommand{\thefigure}{B.\arabic{figure}}
\setcounter{figure}{0}
\begin{minipage}{\textwidth}
    \centering
    \begin{minipage}{0.475\textwidth}
        \centering
        \includegraphics[height=6cm,angle=0.0 ]{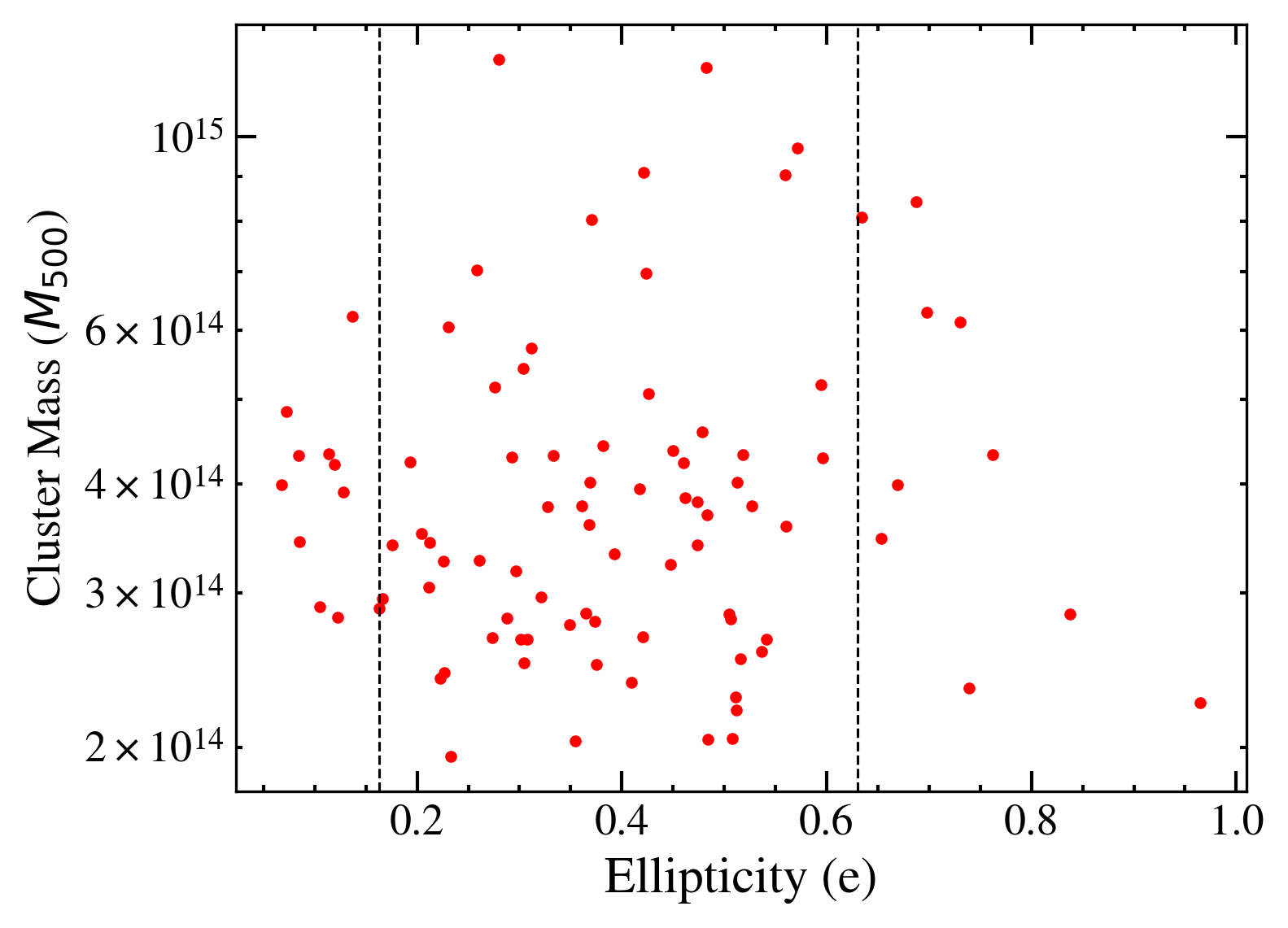}
    \end{minipage}\hfill
    \begin{minipage}{0.475\textwidth}
        \centering
        \includegraphics[height=6cm,angle=0.0 ]{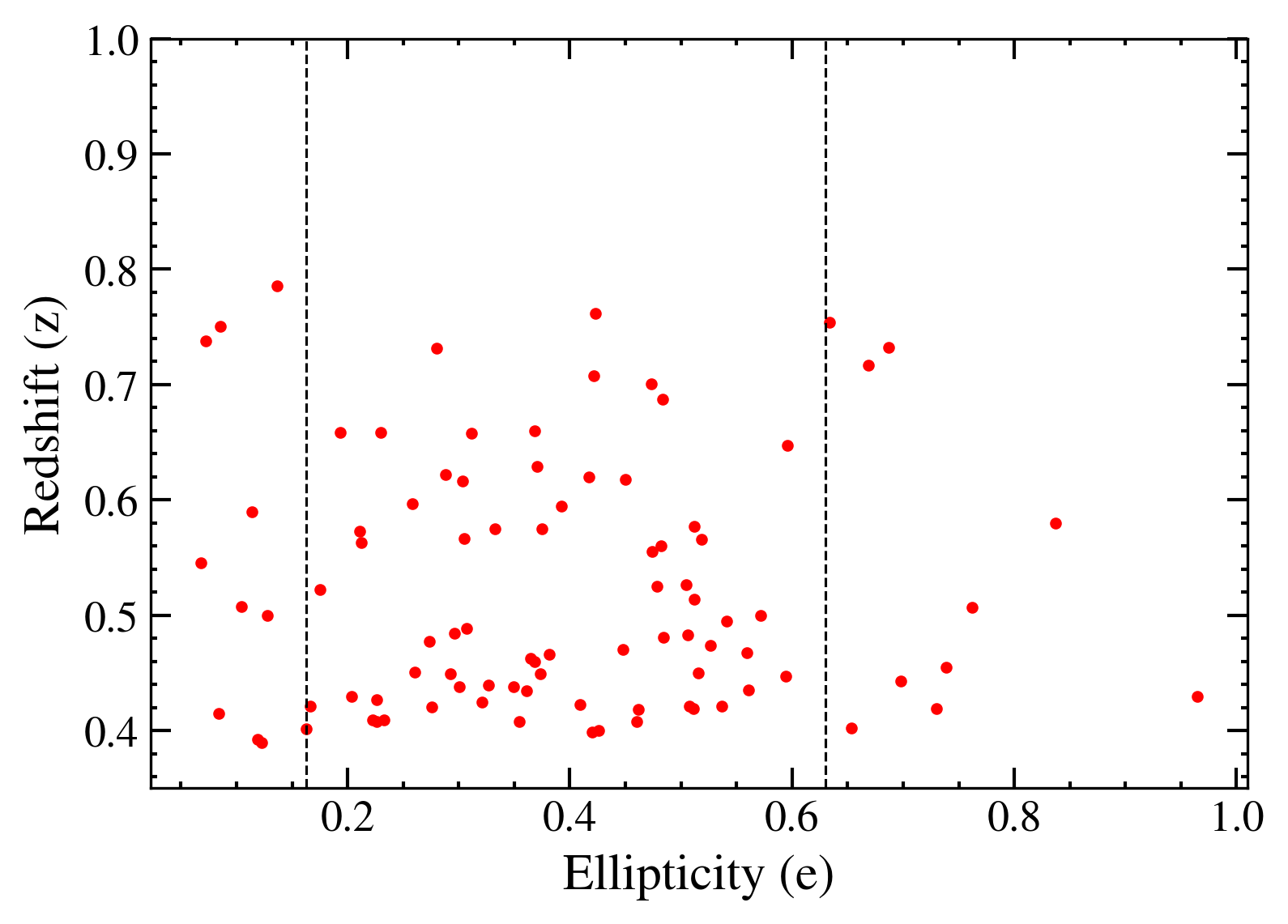}
    \end{minipage}

    \captionof{figure}{Left-hand panel: Total mass ($M_{500}$) versus ellipticity distribution of 91 eFEDS galaxy clusters from our sample of 105 clusters for which the ellipticities are given by \cite{Klein2022A&A...661A...4K}. Right-hand panel: Redshift versus ellipticity distribution of the same galaxy cluster sample. The vertical lines at ellipticity values of 0.16 and 0.63 separate the 10 lowest and the 10 highest ellipticity clusters from our cluster sample.}
    \label{fig:ellip_mass_redshift_distribution}
\end{minipage}

\end{document}